\documentclass{article} 
\pdfoutput=1 
\newcommand{\ShoLong}[2]{#2} 

\usepackage[ansinew]{inputenc}
\usepackage[dvips]{graphicx}
\usepackage{graphics}
\ShoLong{}{\usepackage{amsthm}}
\usepackage{amssymb,amsfonts,amsmath}
\usepackage {color}

\newcommand{\Poly}{\ensuremath{\mathcal{P}}}

\newcommand{\CH}{\ensuremath{\mathcal{CH}}}
\newcommand{\R}{\ensuremath{\mathbb{R}}}
\newcommand{\seq}[1]{\ensuremath{\left \langle #1 \right \rangle}}

\graphicspath{{figs/}}

\newcommand{\ray}[2]{\ensuremath{r(#1,#2)}}

\title{Geodesic-Preserving Polygon Simplification\thanks{%
Research supported by the ESF EUROCORES
programme EuroGIGA - ComPoSe, Austrian Science Fund (FWF): I
648-N18 and grant EUI-EURC-2011-4306.
T.H.\ supported by the Austrian Science Fund (FWF): P23629-N18 `Combinatorial Problems on Geometric Graphs'.
M.K.\ received support of the Secretary for Universities and Research of the Ministry of Economy and Knowledge of the Government of Catalonia and the European Union.
A.P.\ is recipient of a DOC-fellowship of the Austrian
Academy of Sciences at the Institute for Software Technology, Graz University
of Technology, Austria.
}
}

\author{Oswin Aichholzer\ShoLong{\inst{1}}{\thanks{Institute for Software Technology, Graz University of Technology, Austria, {\tt [oaich|thackl|apilz|bvogt]@ist.tugraz.at}.}}
\and Thomas~Hackl\ShoLong{\inst{1}}{\footnotemark[2]}
\and Matias~Korman\ShoLong{\inst{2}}{\thanks{Dept.\ Matem\`atica Aplicada II, Universitat Polit\`ecnica de Catalunya, Spain, {\tt matias.korman@upc.edu}.}}
\and Alexander~Pilz\ShoLong{\inst{1}}{\footnotemark[2]}
\and Birgit~Vogtenhuber\ShoLong{\inst{1}}{\footnotemark[2]}
}

\ShoLong{\institute{
Institute for Software Technology,
Graz University of Technology,
Austria,
\email{[oaich|thackl|apilz|bvogt]@ist.tugraz.at}.
\and
Dept.\ Matem\`atica Aplicada II, Universitat Polit\`ecnica de Catalunya, 
Spain,
\email{matias.korman@upc.edu}.
}}{}

\usepackage[mathlines]{lineno}
\newcommand*\patchAmsMathEnvironmentForLineno[1]{%
  \expandafter\let\csname old#1\expandafter\endcsname\csname #1\endcsname
  \expandafter\let\csname oldend#1\expandafter\endcsname\csname end#1\endcsname
  \renewenvironment{#1}%
     {\linenomath\csname old#1\endcsname}%
     {\csname oldend#1\endcsname\endlinenomath}}%
\newcommand*\patchBothAmsMathEnvironmentsForLineno[1]{%
  \patchAmsMathEnvironmentForLineno{#1}%
  \patchAmsMathEnvironmentForLineno{#1*}}%
\AtBeginDocument{%
\patchBothAmsMathEnvironmentsForLineno{equation}%
\patchBothAmsMathEnvironmentsForLineno{align}%
\patchBothAmsMathEnvironmentsForLineno{flalign}%
\patchBothAmsMathEnvironmentsForLineno{alignat}%
\patchBothAmsMathEnvironmentsForLineno{gather}%
\patchBothAmsMathEnvironmentsForLineno{multline}%
}
\ShoLong{\let\doendproof\endproof
\renewcommand\endproof{~\hfill$\qed$\doendproof}}{
\newtheorem{theorem}{Theorem}
\newtheorem{observation}{Observation}
\newtheorem{corollary}{Corollary}
\newtheorem{lemma}{Lemma}
\newtheorem{open}{Open problem}

}

\begin{document}

\maketitle

\ShoLong{\spnewtheorem{open}{Open Problem}{\bfseries}{\itshape}
\spnewtheorem{observation}{Observation}{\bfseries}{\itshape}}{}

\begin{abstract}
Polygons are a paramount data structure in computational geometry.
While the complexity of many algorithms on simple polygons or polygons
with holes depends on the size of the input polygon, the intrinsic
complexity of the problems these algorithms solve is often related to the reflex vertices of the polygon.
In this paper, we give an easy-to-describe linear-time method to replace an input polygon~$\Poly$ by a polygon $\Poly'$ such that (1)~$\Poly'$ contains~$\Poly$, (2)~$\Poly'$ has its reflex vertices at the same positions as~$\Poly$, and (3)~the number of vertices of $\Poly'$ is linear in the number of reflex vertices.
Since the solutions of numerous problems on polygons (including shortest paths, geodesic hulls, separating point sets, and Voronoi diagrams) are equivalent for both $\Poly$ and $\Poly'$, our algorithm can be used as a preprocessing step for several algorithms and makes their running time dependent on the number of reflex vertices rather than on the size of~$\Poly$.
\ShoLong{}
{We describe several of these applications (including linear-time post-processing steps that might be necessary).}
\end{abstract}

\section{Introduction}
\ShoLong{}{
\subsection{Definitions and Results}%
}
A \emph{simple polygon} is a closed connected domain in the plane that is bounded by a sequence of straight line segments (edges) such that any two edges may intersect only in their endpoints (vertices), and such that in every vertex exactly two edges intersect.
Let $\Poly$ be a simple polygon and let $\mathcal{H}_1,\ldots,\mathcal{H}_k$ be a set of pairwise-disjoint simple polygons such that $\mathcal{H}_i$ is contained in the interior of~$\Poly$ for $1 \leq i \leq k$.
Then the closure~$\mathcal{Q}$ of $\Poly \setminus \bigcup_{1 \leq i \leq k} \mathcal{H}_i$ is called a \emph{polygon with holes}.
The polygons~$\mathcal{H}_i$ are called the \emph{holes} of~$\mathcal{Q}$, and the vertices and edges of $\mathcal{Q}$ are the vertices and edges of $\Poly$ and all $\mathcal{H}_i$, respectively.
We regard $\Poly$ and $\mathcal{Q}$ as closed sets, i.e., they include their boundary.
All polygons we consider are either simple polygons or polygons with holes.
A vertex of a polygon is reflex if its inner angle is larger than $180^\circ$.
Given a polygon $\Poly$ with $n$ vertices, the \emph{geodesic path} $\pi_{\Poly}(p,q)$ between two points $p$ and $q$ of $\Poly$ is defined as the shortest path that connects $p$ and $q$ among all the paths that stay within $\Poly$.
The length of that path is called the \emph{geodesic distance}.
For any pair of points in~$\Poly$, such a path always exists and, when $\Poly$ is simple, is unique\ShoLong{}{ (in contrast to polygons with holes)}.
Moreover, such a path is a polygonal chain whose vertices (other than $p$ and $q$) are reflex vertices of $\Poly$.
When the path $\pi_{\Poly}(p,q)$ is a straight line segment, we say that $p$ \emph{sees} $q$ (and vice versa).
\ShoLong{}{We regard both polygons and geodesics as closed subsets of the plane.}

\begin{figure}[htb]
\centering
\includegraphics[page=2]{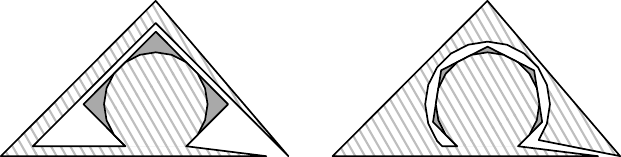}
\caption{Simple polygons (dashed) drawn on top of subsuming polygons (solid).}
\label{fig_example1}
\end{figure}

\ShoLong{We say that two polygons $\Poly$ and $\Poly'$ have the same reflex vertices if for any reflex vertex $v\in \Poly$ there is a reflex vertex $v'\in \Poly'$ at the same point in the plane and vice versa.}{%
We say that two polygons $\Poly$ and $\Poly'$ have the same reflex vertices if any reflex vertex $v\in \Poly$ is at the same point in the plane as a vertex of~$\Poly'$ and $v$ is also reflex in~$\Poly'$ (analogously, any reflex vertex of $\Poly'$ must also be a reflex vertex of~$\Poly$).} 
We say that $\Poly'$ \emph{subsumes} $\Poly$ if $\Poly$ and $\Poly'$ have the same reflex vertices and $P\subseteq P'$.
See \figurename~\ref{fig_example1} for examples.
\ShoLong{}{For subsuming polygons, we can make the following observation.}

\begin{observation}
Let $\Poly, \Poly'$ be two simple polygons such that $\Poly'$ subsumes $\Poly$.
Then, for any $p,q\in \Poly$ we have $\pi_{\Poly}(p,q)=\pi_{\Poly'}(p,q)$.
\end{observation}

For algorithms that solely rely on geodesic paths inside $\Poly$, the output will remain equivalent if we replace $\Poly$ by $\Poly'$ in the input.
It is therefore desirable to construct a subsuming polygon $\Poly'$ such that (1)~$\Poly'$ has few vertices and (2)~the construction can be done in time linear in the size of~$\Poly$.
For the creation of a subsuming polygon of minimal size, one has to connect the reflex vertices by pairwise non-intersecting paths in the union of the exterior and the boundary of the original polygon.
Guibas, Hershberger, Mitchell, and Snoeyink~\cite{approximating_polygons} address various aspects of the problem of approximating polygonal paths and polygons by simpler ones.
They show that the problem of finding a minimum link simple polygon (of a given homotopy class) having its boundary inside a given region~$R$ is NP-hard.
In this paper, we show the following result.
\begin{theorem}\label{theo_main}
For any polygon $\Poly$ (possibly with holes) of $n$ vertices out of which $r > 0$ are reflex, there exists a polygon $\Poly'$ with $O(r)$ vertices that subsumes $\Poly$. Moreover, $\Poly'$ can be computed in $O(n)$ time and will have the same number of holes as $\Poly$.
\end{theorem}

Before giving the proof of Theorem~\ref{theo_main} in Section~\ref{sec_proof}, we review some basic properties of pointed pseudo-triangulations (also known as \emph{geodesic triangulations}) in Section~\ref{sec_pt}.
Section~\ref{sec_applications} gives an overview of the various applications of our result.
In Section~\ref{sec_conclusion} we summarize our approach and give a short account on the open problem of computing an optimal subsuming polygon. \ShoLong{Due to space constraints, the proofs of some claims are deferred to the full version of the paper.}{}

\ShoLong{
\subsection*{Related Work}
}
{
\subsection{Related Work}
}
We follow the common aim of relating the time and space complexity of algorithms on polygons not only to the input size, but also to the number of reflex vertices.
This is often a more significant parameter for the actual difficulty of the problem instance.
Hertel and Mehlhorn~\cite{hertel_mehlhorn} give an algorithm for triangulating a simple polygon of $n$ vertices, $r$ of which are reflex, in~$O(n \log r)$ time.%
\ShoLong{}{\footnote{While the $O(n)$ time algorithm of Chazelle~\cite{chazelle} is asymptotically better, this algorithm is considered as being difficult to implement (see, e.g.,~\cite[p.~57]{o_rourke_c}).
Therefore, theoretically suboptimal algorithms for that problem are still relevant.}}
Bose et al.~\cite{bdhilm-ghsc-04} give a~$\Theta(m+n \log r)$ algorithm for computing a geodesic ham-sandwich cut of two given sets of $m$ points in a simple polygon with $r$ reflex vertices.
\ShoLong{}{Keil~\cite{keil} describes various decomposition algorithms that are dependent on the number of reflex vertices.
The number of reflex vertices in a simple polygon has also been used as a parameter for combinatorial problems, see, e.g.,~\cite{bchm-gwt-11,hurtado_noy_vis_graph}.}

A related way of giving a more fine-grained analysis of algorithms on polygons is by expressing its complexity in the number of edges in the visibility graph (i.e., the number of point pairs that see each other), see for example\ShoLong{}{~\cite{bern_eppstein} and}~\cite[p.~68]{ghosh}.
Note that, e.g., for computing the minimum weight triangulation of a simple polygon, the currently known worst case appears when there are no reflex vertices~\cite{bern_eppstein}.

The term ``polygon simplification'' is also used in connection with operations that are used to compress polygons and to reduce noise in the representation.
A well-known algorithm to smoothen polygonal chains is the Douglas-Peucker algorithm~\cite{douglas_peucker}.
Guibas et al.~\cite{approximating_polygons} address several variations of the approach to fatten existing polygonal chains and approximate the chain inside the fattened region.
There also exists work on constructing simple polygons that contain given ones and fulfill certain properties, as a generalization of the convex hull of simple polygons~\cite{minimum_area_hulls};
the main objective there is to approximate the shape.

\section{Pseudo-Triangulations}\label{sec_pt}
\ShoLong{In this section, we recall several properties of pseudo-triangles and pointed pseudo-triangulations of simple polygons.
For details on pseudo-triangulations in various contexts see the survey by Rote, Santos, and Streinu~\cite{pt_survey}.}{%
In this section, we recall several properties of pseudo-triangles and pointed pseudo-triangulations that will be used throughout the rest of the paper.
We exclusively consider pointed pseudo-triangulations of simple polygons.
For more details on pseudo-triangulations see the survey by Rote, Santos, and Streinu~\cite{pt_survey}.
}

A \emph{pseudo-triangle} is a simple polygon with exactly three convex vertices.
The three convex vertices are the \emph{corners} of the pseudo-triangle, and the three polygonal chains between the corners are called the \emph{side chains}.
Note that a side chain might consist only of one edge.

A \emph{pseudo-triangulation} of a simple polygon $P$ is a partition of $P$ into pseudo-triangles, such that the union of the vertices of the pseudo-triangles is exactly the vertex set of $P$.
A vertex~$v$ of a pseudo-triangulation is called \emph{pointed} if it is incident to a face in which the angle at $v$ is larger than $180^\circ$.
In a \emph{pointed pseudo-triangulation} every vertex is pointed.
Throughout this work, we are only concerned with pointed pseudo-triangulations.
It can be shown that a pointed pseudo-triangulation of a simple polygon with $c$ convex vertices has $c-2$ pseudo-triangles and adds $c-3$ diagonals to the polygon (see, e.g.,~\cite{pt_survey}).
Our main result heavily relies on that fact.
Guibas et al.~\cite{shortest_path_tree} showed that, given a vertex~$v$ of a triangulated simple polygon~$P$, the set of all shortest paths between $v$ and the vertices of~$P$ (i.e., the shortest path tree of~$v$) can be constructed in linear time (this algorithm was later simplified by Hershberger and Snoeyink~\cite{hershberger_snoeyink}).
The union of all these shortest paths gives a pointed pseudo-triangulation of~$P$.
Hence, given a triangulation of a simple polygon~$P$, a pointed pseudo-triangulation of $P$ can be constructed in linear time.%
\footnote{The connection between pointed pseudo-triangulations and the shortest path tree was mentioned by Speckmann and T\'oth~\cite{pi_guards}; the concept of pseudo-triangulations has been developed after the writing of~\cite{shortest_path_tree}.}

Let $\nabla$ be a pseudo-triangle. The line~$\ell$ bisecting the angle at any corner of~$\nabla$ separates the two adjacent side chains $C_1$ and $C_2$, and will leave~$\nabla$ through the third chain~$C_3$.
The bisecting line~$\ell'$ of the angle of a second corner, say, the one joining $C_1$ and $C_3$, separates $C_1$ and $C_3$.
Therefore, $\ell$ and $\ell'$ intersect inside~$\nabla$ and hence separate $C_1$ from $C_2$ and $C_3$%
\ShoLong{.}{%
; see \figurename~\ref{fig_wedge}.
}
This allows us to make the following basic observation.

\ShoLong{}{
\begin{figure}[htb]
\centering
\includegraphics[page=1]{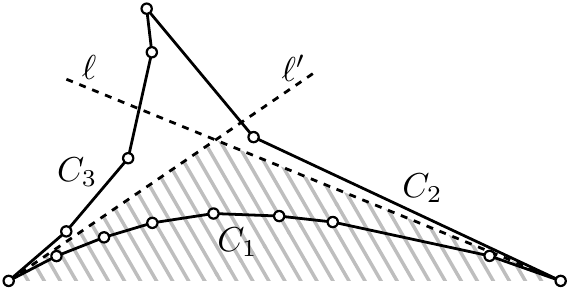}
\caption{The separating wedge of $C_1$.}
\label{fig_wedge}
\end{figure}
}

\begin{observation}\label{obs_separating_wedge}
For any side chain $C$ of a pseudo-triangle~$\nabla$, the bisectors of the angles at the corners of~$C$ define a wedge that separates $C$ from the remaining boundary of~$\nabla$.
\end{observation}

We call this wedge the \emph{separating wedge} of the side chain.
Observation~\ref{obs_separating_wedge} is the crucial property of pseudo-triangles that we will use in the next section.

\section{Proof of the Main Theorem}\label{sec_proof}
\begin{figure}[htb]
\centering
\includegraphics{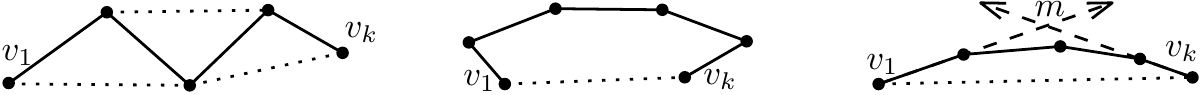}
\caption{A chain that is not hull-honest (left), a hull-honest chain (center), and a simplifiable chain (right).}
\label{fig_monotone_simplifiable}
\end{figure}

We call a simple polygonal chain~$C$ \emph{hull-honest} if it is completely contained in the boundary of its convex hull~$\CH(C)$.
We call a simple chain~$C = \seq{v_1, v_2,\ldots,v_k}$ \emph{simplifiable}, if it is hull-honest and if the ray $\ray{v_1}{v_2}$ (i.e., the ray from $v_1$ through~$v_2$) and the ray $\ray{v_k}{v_{k-1}}$ intersect.
See \figurename~\ref{fig_monotone_simplifiable}.
Note that if a chain is simplifiable, then any of its subchains is simplifiable as well.
For any simplifiable chain $C$ of vertices $v_1,v_2,\ldots v_k$ we introduce an operation called the \emph{simplification} of~$C$ as follows:
If $k \leq 3$ then $C$ remains unchanged.
Otherwise, consider the rays $\ray{v_1}{v_2}$ and $\ray{v_k}{v_{k-1}}$.
As $k > 3$, these two rays intersect at a point $m$ not on the chain.
In that case, we replace $C$ by the chain $C' = \seq{v_1,m,v_k}$.

%
The basic idea of our construction is to simplify long convex chains along the boundary of~$\Poly$.
The main challenge is to avoid that the edges introduced by the simplification intersect with the remaining boundary of the resulting polygon.

\begin{lemma}\label{lem_simpli}
All the side chains of any pseudo-triangle are simplifiable.
When simplifying an arbitrary number of pairwise interior-disjoint subchains, the resulting polygon is again a pseudo-triangle, and, in particular, its boundary is not self-intersecting.
\end{lemma}
\newcommand{\prfsimpli}{
The fact that a side chain of a pseudo-triangle~$\nabla$ is simplifiable follows directly from Observation~\ref{obs_separating_wedge}.
The two rays witnessing simplifiability meet at a point inside the separating wedge of the side chain.
Thus, when simplifying a side chain or one of its subchains, the new chain is completely contained in the intersection of~$\nabla$ and the separating wedge of the original chain.
Further, interior-disjoint subchains of a side chain share at most one vertex, and the same holds for their simplifications; see \figurename~\ref{fig_simpli}.
As 
the corners and their incident angles are not altered by any such simplification, the resulting polygon is again a pseudo-triangle.
\begin{figure}[htb]
\centering
\includegraphics[page=2]{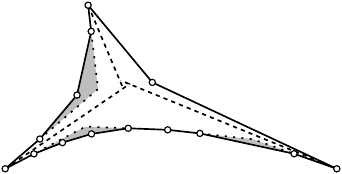}\hfill\hfill
\includegraphics[page=3]{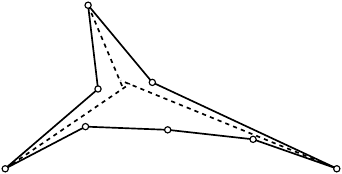}
\caption{A pseudo-triangle and a possible simplification (removing the gray regions).
The separating wedges are indicated by dashed lines.}
\label{fig_simpli}
\end{figure}
}
\ShoLong{
}{\prfsimpli}
Using this result we can now prove our main theorem.

\begin{proof}[Theorem~\ref{theo_main}]
We first consider the case where~$\Poly$ is a simple polygon with at least one reflex vertex.
Let $\CH(\Poly)$ be its convex hull (which can be computed in linear time using, e.g., Melkman's algorithm~\cite{melkman}).
The set $\CH(\Poly) \setminus \Poly$ is the union of (the interiors of) simple polygons whose interiors are pairwise disjoint.
We call these polygons the \emph{pockets}.
Each pocket $P_i$ is defined by exactly one convex hull edge, which is not part of~$\Poly$ and is called a \emph{lid edge}, and a subchain~$C$ of the boundary of~$\Poly$.
The $c_i$ convex vertices of a pocket therefore consist of the reflex vertices of $\Poly$ along $C$ and the two vertices of the lid edge.
Thus, a pointed pseudo-triangulation of $P_i$ has $c_i - 3$ diagonals.
We call the lid edges and the pseudo-triangulation diagonals the \emph{support edges}.
For $p$ pockets, the number of support edges is
\ShoLong{$p + \sum_{i=1}^p (c_i - 3) = p + r + 2p - 3p = r$.
}{
\[
p + \sum_{i=1}^p (c_i - 3) = p + r + 2p - 3p = r \enspace .
\]
}%
Since the only vertices possibly shared by two pockets are the convex hull vertices, we can construct a pointed pseudo-triangulation of each pocket in accumulated $O(n)$ time for all pockets.
See \figurename~\ref{fig_pocketexample} for an example of a pseudo-triangulated pocket.

\begin{figure}[htb]
\centering
\includegraphics{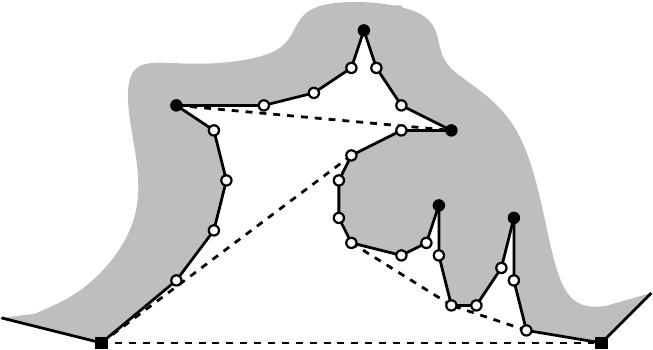}
\caption{A pocket and its pointed pseudo-triangulation.
Support edges are drawn dashed.}
\label{fig_pocketexample}
\end{figure}


Consider a pocket $P_i$ and a pointed pseudo-triangulation~$T(P_i)$ of $P_i$. A side chain of a pseudo-triangle of~$T(P_i)$ consists of convex chains of $\Poly$, possibly separated by support edges. Note that any vertex of $\Poly$ that is not on $\CH(\Poly)$ is clearly part of a side chain of at least one pseudo-triangle of~$T(P_i)$ in some pocket~$P_i$ of~$\Poly$. For each pseudo-triangle in every pocket of~$\Poly$ we simplify all maximal subchains of its side chains that do not contain support edges. Due to Lemma~\ref{lem_simpli}, the resulting polygon is again simple and subsumes~$\Poly$.


Observe that $\CH(\Poly)$ consists of convex chains of $\Poly$, possibly separated by lid edges. Unlike the maximal convex chains inside pseudo-triangles, there might exist a maximal convex chain (not containing lid edges) $C = \seq{v_i, \ldots, v_j}$ on $\CH(\Poly)$ that is not simplifiable. The reason for this being, that the turn of the chain is at least $180^\circ$, i.e., the rays $\ray{v_i}{v_{i+1}}$ and $\ray{v_j}{v_{j-1}}$ do not intersect. However, such a chain $C$ appears at most once on~$\Poly$ and can be split into two simplifiable parts with a common vertex $v^*$.

It remains to count the number of convex vertices of the resulting polygon~$\Poly'$.
To this end, we charge the convex vertices of~$\Poly'$ either to reflex vertices or to one of the $r$ support edges.
%
%
Let $C = \seq{v_i,\ldots,v_j}$ be a maximal convex chain not containing support edges of $\Poly'$.
Each of the two end points of $C$ is either a reflex vertex of $\Poly$ (or $v^*$), or end point of a support edge, or both. If $v_i$ is the end point of a support edge $e$, then we charge $C$ to $e$. Otherwise, we charge $C$ to the reflex point $v_i$ (or to the special point $v^*$). Observe that each end point of a support edge can be at most once a starting point ($v_i$) of such a maximal convex chain $C$ of $\Poly'$. The same is true for each reflex vertex of $\Poly$ (and $v^*$). Thus, $\Poly'$ consists of at most $2r+r+1$ maximal convex chains.

Each such maximal convex chain of $\Poly'$ consists of at most three vertices that might all be convex in $\Poly'$. Observe though, that the last vertex $v_j$ of a chain $C$ is either a reflex vertex or also the first vertex of another maximal convex chain $C'=\seq{v_j,\ldots,v_k}$ of $\Poly'$. Hence, we need to count only at most two convex vertices per chain.
Therefore, $\Poly'$ has at most $6r+2$ convex vertices.

%
%

Finally, observe that we can apply the same strategy to a polygon~$\Poly$ with holes~$\mathcal{H}_i$ by considering each hole as a simple polygon itself.
Recall that our approach simplifies a polygon, preserving geodesics inside the polygon.
We independently pseudo-triangulate the interior of each hole~$\mathcal{H}_i$ (in time linear in the number of vertices of~$\mathcal{H}_i$) and apply the simplification strategy.
For each~$\mathcal{H}_i$ we obtain a polygon whose complexity is proportional to the number of convex vertices of (the polygon) $\mathcal{H}_i$ and preserves geodesics outside~$\mathcal{H}_i$ (i.e., paths that consider~$\mathcal{H}_i$ as an obstacle).
Thus, the resulting simplification preserves geodesics inside~$\Poly$.
As each convex vertex of~$\mathcal{H}_i$ is a reflex vertex of~$\Poly$, the complexity of this simplification will be proportional to the number of vertices that are reflex in~$\Poly$.
\end{proof}

\paragraph{Remark.}
Note that with this process we can explicitly give the coordinates of~$\Poly'$.
If the coordinates of the vertices of~$\Poly$ are rational, then the coordinates used for~$\Poly'$ are rational as well. 
Moreover, explicitly representing these vertices needs at most a constant times the number of bits used by input vertices.   
Alternatively, we can store the simplification in an implicit form: we store the first and last vertices of each simplified chain, allowing an easy identification between each connected component of $\Poly' \setminus \Poly$ and the simplified chains.
In either case, at most $O(r)$ space is needed (for constant size vertex representation).

While our result is asymptotically optimal, we do not construct a subsuming polygon with the minimal number of vertices.
Given a polygon~$\mathcal{Q}$ with $h$ holes, Guibas et al.~\cite[Theorem~5]{approximating_polygons} show how to construct a simple polygon~$\Poly$ with its boundary inside~$\mathcal{Q}$ such that $\Poly$ contains all holes and has only $O(h)$ more vertices than the (probably non-simple) optimum.
We note that, after knowing that the number of vertices of a minimal subsuming polygon is in~$O(r)$, also the method by Guibas et al., after careful modification, could be used for constructing a subsuming polygon within the given bounds when applying it to each pocket.
However, our constructive proof of the bound gives a straight-forward algorithm that can be implemented using standard tools used for visibility problems in simple polygons.

\section{Applications}\label{sec_applications}
Let $\mathcal{A}$ be an algorithm that receives a polygon $\Poly$ (and potentially other input~$I$). We say that $\mathcal{A}$ is \emph{subsuming} if the result of executing $\mathcal{A}$ with input~$\Poly$ and~$I$ is equal to the result obtained when the input is $\Poly'$ and $I$, where $\Poly'$ is a polygon that subsumes $\Poly$. Note that, if $\mathcal{A}$ is subsuming, so will any other algorithm that solves the same problem (thus we say that the problem is {\em simplifiable}).

\begin{theorem}
\label{thm_application}
\sloppypar{
Let $\Poly$ be a polygon of size $n$ with $r$ reflex vertices and let~$I$ be additional input of size~$m$. Let $\mathcal{A}(\Poly,I)$ be an algorithm that solves a simplifiable problem and runs in $T(n,m)$ time using $S(n,m)$ space. Then, $\mathcal{A}(\Poly,I)$ can be modified to run in $O(n+T(r,m))$ time and $O(n+S(r,m))$ space. 
}
\end{theorem}

In the following we present several applications for Theorem~\ref{thm_application} which show the versatility of our approach.
Most of these problems can be stated in terms of shortest paths, which immediately implies that the problems are simplifiable. 
For each task, we briefly state the problem, mention the (to the best of our knowledge) fastest existing algorithm, and explain how our approach helps to reduce the running time.

\subsection{Shortest Paths}
Computing the shortest path that avoids a series of obstacles and connects two given points belongs to the most fundamental problems in computational geometry.
When looking for the shortest path between two given points inside a simple polygon~$\Poly$ of size $n$, the currently best known algorithm is due to Guibas and Hershberger~\cite{gh-ospqsp-89}: they provide a method that, after an $O(n)$ time preprocessing, can report the geodesic distance between any two points in~$\Poly$ in $O(\log n)$ time.
If the geodesic is to be reported, their method needs $O(k+\log n)$ time instead, where $k$ is the number of vertices of the path. 
By applying our polygon simplification strategy, we reduce the query time to $O(\log r)$.

\begin{corollary}
We can preprocess a simple polygon $\Poly$ of $n$ vertices out of which $r>0$ are reflex in $O(n)$ time and space, such that for any two points $p,q\in \Poly$ we can determine their geodesic distance $|\pi_{\Poly}(p,q)|$ in $O(\log r)$ time.
Moreover, the geodesic path can be reported in $O(k+\log r)$ time, where $k$ is the number of vertices of $\pi_{\Poly}(p,q)$. 
\end{corollary}

\sloppypar{
Many variations of the above problem have been studied in the
literature (see~\cite{m-spn-04} for a survey). Among several results, we highlight the algorithm of Hershberger and Suri~\cite{hs-oaespp-99} for computing shortest paths in the case where holes are also present. The running time of their preprocessing algorithm is bounded by $O(n\log n)$, which can again be improved by our approach.
}

\begin{corollary}
We can preprocess a polygon $\Poly$ with holes of $n$ vertices out of which $r>0$ are reflex in $O(n+r\log r)$ time and space, such that for any point $p\in \Poly$ we can determine the geodesic distance $|\pi_{\Poly}(p,q)|$ with respect to a fixed point $q\in \Poly$ in $O(\log r)$ time.
Moreover, a geodesic path can be reported in $O(k+\log r)$ time, where $k$ is the number of vertices of $\pi_{\Poly}(p,q)$. 
\end{corollary}

Shortest path computation has also been studied in other metrics, like, for example, the $L_1$ metric~\cite{m-l1spapop-92}. Although the proposed algorithm claims a running time of $O(n\log n)$, it is easy to see that convex vertices only contribute a linear fraction to the running time. Thus, the running time of the algorithm in~\cite{m-l1spapop-92} can be bounded by $O(n+r\log r)$ using a simple counting argument.

\subsection{Geodesic Hull}
A set $S\subseteq \Poly$ is called {\em geodesically} (or relative) convex if and only if for any $p,q\in S$, it holds that their geodesic $\pi_\Poly(p,q)$ is in $S$.
The \emph{geodesic hull} of a set $S\subseteq \Poly$ is defined as the (inclusion-wise) smallest geodesically convex set that contains $S$.

Given a set $S$ of $m$ points and a simple polygon $\Poly$ of $n$ vertices, Toussaint~\cite{relativehull} studied the problem of determining whether $S$ is geodesically convex, and\,---\,if not\,---\,computing its geodesic hull. The proposed algorithm runs in $O((n+m)\log{(n+m)})$ time, which we can reduce with our approach.


\begin{corollary}
Given a simple polygon $\Poly$ of $n$ vertices out of which $r>0$ are reflex, we can compute the geodesic hull of a given  set $S$ of $m$ points in the interior of $\Poly$ in $O(n+(m+r)\log (m+r))$ time using $O(n+m)$ space. 
\end{corollary}

\subsection{Separating Point Sets}
Given two sets $R$ and $B$ of $m$ points in $\R^2$ and a geometric object $\zeta$ (usually a line) that partitions the plane into two components, we say that $\zeta$ {\em separates} $R$ and $B$ if each component of the plane only contains elements of one of the two sets.
The extension of separability to polygons with holes was studied by Demaine et al.~\cite{dehilmow-spspe-05}.
They showed that finding the minimum number of chords that separate the two sets inside a polygon with holes is NP-hard. 

However, if $\Poly$ is a simple polygon, the problem becomes easier: in~\cite{dehilmow-spspe-05} the authors give necessary and sufficient conditions for the existence of a separating geodesic in simple polygons, which results in an algorithm that runs in $O((n+m)\log{(n+m)})$ time to determine the separability of $R$ and $B$. The combination of their result with our simplification technique reduces the dependency in $n$.

\begin{corollary}
Given a simple polygon $\Poly$ of $n$ vertices out of which $r>0$ are reflex, and two sets $R$  and $B$ of $m$ points each, we can determine whether or not $R$ and $B$ are separable by a geodesic (and find a separating geodesic, if any exists) in $O(n+(m+r)\log (m+r))$ time using $O(n+m)$ space. 
\end{corollary}

\subsection{Triple Orientation}
Many algorithms for point sets in the plane only use the orientations of point triples, i.e., a ternary predicate $p(a,b,c)$.
In analogy to unconstrained point sets, the orientation of a point triple inside a simple polygon is defined via the order in which the points appear on the geodesic hull of the triple~\cite{geodesic_order_types}.

\begin{corollary}\label{cor_orientation}
We can preprocess a simple polygon $\Poly$ of $n$ vertices out of which $r>0$ are reflex in $O(n)$ time and space so that, for any three points $a,b, c\in \Poly$, we can determine their orientation in $O(\log r)$ time. 
\end{corollary}
\newcommand{\prforientation}{
Suppose we want to obtain the orientation of the point triple $abc$.
We use the method of~\cite{gh-ospqsp-89} for reporting geodesics inside a triangulated polygon.
We have to consider three different cases.
If (1) all three points are in the same triangle, the orientation is given trivially.
Suppose (2) that $b$ and $c$ are in the same triangle, but $a$ is not.
Then we report the first segment of the geodesics $\pi_\Poly(b,a)$ and $\pi_\Poly(c,a)$, from which the orientation follows.
If (3) all three points are in different triangles, then the orientation of the triple is obtained using the dual graph of the triangulation (which is a tree).
The triple orientation can be derived in the following way by a constant number of ancestor queries in the dual tree, which are doable in constant time, after preprocessing the tree (once) in linear time (see~\cite[Sect.~3]{gh-ospqsp-89}).
Let $\Delta_a, \Delta_b$, and $\Delta_c$ be the triangles that contain $a$, $b$, and $c$, respectively.
If (3.1) according to these queries w.l.o.g.\ $\Delta_b$ is on a path between $\Delta_a$ and $\Delta_c$, then we report the first segment of the geodesics $\pi_\Poly(b,a)$ and $\pi_\Poly(b,c)$, which give the orientation at $b$.
Otherwise (3.2), the order in which $\Delta_a$, $\Delta_b$, and $\Delta_c$ are traversed by a depth-first traversal of the tree (which is done during preprocessing), directly gives their orientation.
}

\ShoLong{
}{
\begin{proof}
\prforientation
\end{proof}
}
In~\cite{geodesic_order_types}, it was shown that each point set in a simple polygon corresponds to an \emph{abstract order type}, a generalization of point set order types in the plane (roughly speaking, an order type is an equivalence class of point sets w.r.t.\ the predicate~$p$).
Algorithms that operate on abstract order types can be applied in this setting, using $O(\log r)$ time per orientation test.
For example, using the results of~\cite{extreme_point} one can compute a halving geodesic through a given point of a set~$S$ of $m$ points, and also the geodesic hull edges stabbed by (the extension of) a geodesic through two given points in $O(n+m\log r)$ time.

\subsection{Voronoi Diagram}
Voronoi diagrams are another fundamental data structure in computational geometry, hence, it is no surprise that the geodesic variant was studied in the late 80s~\cite{a-ogvdpssp-87}. Since then the algorithms have been improved, and the currently best algorithm is due to Papadopoulou and Lee~\cite{pl-anagvdpsporpd-98}. 
The furthest-site Voronoi diagram has also been studied in geodesic environments~\cite{afw-fsgvd-93}. 
The fastest algorithms for computing either diagram run in $O((n+m)\log{(n+m)})$ time, and use $O(n+m)$ space~\cite{afw-fsgvd-93,pl-anagvdpsporpd-98}. 
It is easy to see that, for any point $q\in\Poly$, its nearest or furthest site (w.r.t.\ geodesic distance) will be the same in $\Poly$ and in any polygon that subsumes $\Poly$. 
Thus, in principle our approach can be used. Since the boundary is part of the Voronoi diagram, some post-processing will be necessary for this problem.

\newcommand{\prfvoronoi}
{%
The nearest- and furthest-site geodesic Voronoi diagram both are a concatenation of $O(n+m)$ straight and hyperbolic arcs, proven by Aronov~\cite{a-ogvdpssp-87} and Aronov et~al.~\cite{afw-fsgvd-93}, respectively.

By the ``Ordering Lemma''~\cite{afw-fsgvd-93} for furthest-site geodesic Voronoi diagrams, the ordering of the sites with nonempty Voronoi cells around the geodesic hull of $S$ is the same as the ordering of Voronoi cells around the boundary of $\Poly$. As the geodesic hull of $S$ with respect to $\Poly$ stays unchanged for $\Poly'$, also this ordering of Voronoi cells is the same on the boundary of $\Poly'$. Thus, the union of the arcs of the furthest-site geodesic Voronoi diagram in~$\Poly'$ is a superset of the union of the arcs of~$\Poly$, and the connected components of the furthest-site geodesic Voronoi diagram in $\Poly' \setminus \Poly$ are paths.

Observe that there is no site in $\Poly' \setminus \Poly$. Therefore, in a nearest-site geodesic Voronoi diagram there cannot be a complete Voronoi cell in $\Poly' \setminus \Poly$, and the edge graph of the nearest-site geodesic Voronoi diagram inside $\Poly' \setminus \Poly$ is cycle-free. Further, the union of the arcs of the nearest-site geodesic Voronoi diagram in~$\Poly'$ is a superset of the union of the arcs of~$\Poly$.

For both versions of a geodesic Voronoi diagram, it remains to obtain the exact points where the geodesic Voronoi diagram and the boundary of~$\Poly$ intersect. The basic idea is to traverse the boundary of $\Poly$ and to obtain the intersections with the arcs. This is straight-forward if $\Poly$ and $\Poly'$ are intersected by the same arc. However, the part of the geodesic Voronoi diagram in a connected component of $\Poly' \setminus \Poly$ might be a forest.

Suppose that during our traversal of $\Poly$ we reach a vertex $v_i$ where a simplification $C' = \seq{v_i, m, v_j}$ of a convex chain $C = \seq{v_i, \ldots, v_j}$ starts.
We obtain the arc $a$ of the current Voronoi cell that intersects $C'$ next (if no such arc exists, there is no intersection with the geodesic Voronoi diagram and $C$, and we continue at~$v_j$).
For nearest-site geodesic Voronoi diagrams Aronov~\cite{a-ogvdpssp-87} shows that the bisecting geodesic of any two points intersects the boundary of the surrounding polygon (i.e., both $\Poly$ and $\Poly'$) exactly twice.
For furthest-site geodesic Voronoi diagrams a similar result is given in~\cite{afw-fsgvd-93} and also follows from the above mentioned ``Ordering Lemma''.
Hence, every arc of the geodesic Voronoi diagram is either entirely on one side of an edge~$e$ of the polygon, or is intersected exactly once by the supporting line of~$e$.
Thus, the arc $a$ is the root of a tree in the connected component of $\Poly' \setminus \Poly$ bounded by $C$ and $C'$. We apply the basic linear-time approach to find the intersection(s) of $C$ with the tree rooted at $a$:
Let $x$ be the arc $a$ and let $e$ be the edge $v_i v_{i+1}$. If $x$ is to the right of (the supporting line of) $e$ (using a counterclockwise representation of the polygons), then we store $x$ on a stack and continue with the next son of $x$ as the new arc $x$. If $x$ is not intersected by $e$, but by the supporting line of $e$, then we continue with $e=v_{i+1} v_{i+2}$. If the edge $e$ intersects $x$, we prune the arc $x$ at this intersection. We use the stack to backtrack to the next unprocessed arc in the tree and continue with this arc as $x$. (Note that using a stack and backtracking is not needed for furthest-site geodesic Voronoi diagrams, as in this case there are no Voronoi vertices in $\Poly' \setminus \Poly$.) If the stack becomes empty during backtracking, we have processed the whole tree for arc $a$ and continue with the next arc intersecting $C'$.
In every step, we either traverse an edge of $\Poly$ or discard one of the $O(n+m)$ arcs of the initial geodesic Voronoi diagram.  Hence, we prune the geodesic Voronoi diagram of~$\Poly'$ to the one of~$\Poly$ in~$O(n+m)$ time.%
}

\begin{corollary}\label{cor_voronoi}
Given a simple polygon $\Poly$ of $n$ vertices out of which $r>0$ are reflex, and a set $S$ of $m$ sites, we can compute the nearest- and furthest-site geodesic Voronoi diagram of $S$ with respect to $\Poly$ in $O(n+(m+r)\log (m+r))$ time using $O(n+m)$ space. 
\end{corollary}
\ShoLong{
}
{
\begin{proof}
\prfvoronoi
\end{proof}
}

\subsection{Geodesic Diameter, Median, and Maximum Distance}
Given a set $S$ of $n$ sites in a simple polygon~$\Poly$, the \emph{geodesic median} is the site of~$S$ that minimizes the maximum geodesic distance to points of $S$. The \emph{geodesic diameter} of $S$ is the maximal distance between any two sites of $S$. If we are given two sets of sites $S_1$ and $S_2$ instead, their \emph{maximum geodesic distance} is defined as the largest distance between points of different sets. Toussaint~\cite{relativehull} introduced these concepts and gave algorithms on how to compute them. The running times of the proposed algorithms range between $O((n+m)\log (n+m))$ and $O(n^2+m^2)$. 
As pointed out by Toussaint, these problems can also be solved by computing the furthest-site geodesic Voronoi diagram and making $O(m)$ queries. If we simplify the polygon before making the queries, we obtain the following result.

\begin{corollary}
Given a simple polygon $\Poly$ of $n$ vertices out of which $r>0$ are reflex, and two sets $S_1$, $S_2$ of $m$ sites, we can compute the geodesic median, the geodesic diameter of $S_i$  (for $i\leq 2$), and the maximum geodesic distance between $S_1$ and $S_2$ in $O(n+(m+r)\log(m+r))$ time using $O(n+m)$ space.  
\end{corollary}

\section{Conclusion}\label{sec_conclusion}

As already mentioned, while asymptotically tight, our method does not construct optimal subsuming polygons, i.e., subsuming polygons with the minimum number of vertices.
In particular, every edge of the subsuming polygon~$\Poly'$ completely contains an edge of the initial polygon~$\Poly$.
One can easily construct examples where this is not the case for optimal subsuming polygons%
\ShoLong{.}
{, see~\figurename~\ref{fig_simplify_no_support}.
\begin{figure}[htb]
\centering
\includegraphics{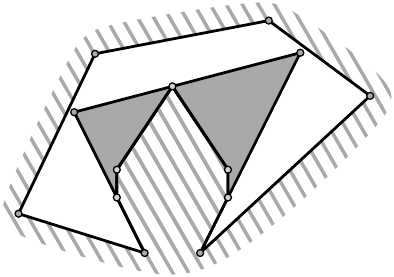}
\caption{A hole in a polygon where an edge of any optimal subsuming polygon does not completely contain an edge of the initial polygon.}
\label{fig_simplify_no_support}
\end{figure}
}%
\ShoLong{}{%
However, one can show that there always is an optimal subsuming polygon~$\Poly'$ such that the supporting line of every edge of $\Poly'$ contains a vertex of $\Poly$.
Note that in the example of \figurename~\ref{fig_simplify_no_support}, the chain added to the subsuming polygon is a minimum link path between the two reflex vertices inside the hole.
We have the additional requirement that the path is convex, but, since the paths are restricted by a convex obstacle, any minimum link path between two consecutive reflex vertices will be convex.}

An optimal subsuming polygon consists of several disjoint paths with a minimum overall number of vertices.
In general, the minimum link path problem is 3SUM-hard~\cite{link_paths_revisited} even for a single path.
However, the reduction used in~\cite{link_paths_revisited} is not applicable to our restricted setting where the paths are in a domain without holes.
Inside simple polygons, the minimum link path problem is solvable in linear time~\cite{shortest_link_path_linear} for a single path.

Guibas et al.~\cite{approximating_polygons} show that finding a minimum link simple polygon having its boundary inside a given region~$R$ is NP-hard, 
but they point out that their reduction requires holes in~$R$ on both sides of the polygon boundary.

The pairwise-disjoint link paths problem inside a simple polygon is discussed by Gupta and Wenger~\cite{gupta}.
They give a constant-factor approximation algorithm, and ask whether there exists a polynomial-time algorithm for computing the optimum.
We state the same question for our restricted setting
\ShoLong{: }{. \begin{open}}%
Given a simple polygon $\Poly$, what is the complexity of constructing a subsuming polygon with the least number of (convex) vertices?
\ShoLong{}{\end{open}}
%

\bibliographystyle{splncs03}
\bibliography{bibliography}

\end{document}